\documentclass[twocolumn,showpacs,preprintnumbers,amsmath,amssymb]{revtex4}

\usepackage{graphicx}% Include figure files
\usepackage{dcolumn}% Align table columns on decimal point
\usepackage{bm}% bold math

\begin{document}

\preprint{APS/123-QED}

\title{A Brownian particle having a fluctuating mass}

\author{R. Lambiotte}
\email{Renaud.Lambiotte@ulg.ac.be}

\author{M. Ausloos}
\email{Marcel.Ausloos@ulg.ac.be}

\affiliation{%
SUPRATECS,  B5 Sart-Tilman, B-4000 Li\`ege, Belgium
}%

\date{05/04/2005}

\begin{abstract}
We focus on the dynamics of a Brownian
particle whose mass fluctuates. First we show that the behaviour is similar
  to that of a Brownian particle moving in a fluctuating medium,
as studied by Beck [{\em Phys. Rev. Lett.} { 87} {(2001)} {180601}].
   By performing numerical simulations of the Langevin equation, we
check the theoretical predictions derived in the adiabatic limit, and
study deviations outside this limit. We compare the
mass velocity distribution with {\em truncated} Tsallis distributions
[{\em  J. Stat. Phys.} {52} {(1988)} {479}] and find excellent
agreement if the masses are chi-squared distributed. We also consider 
the diffusion of the Brownian
particle by studying a Bernoulli random walk with fluctuating walk 
length in one dimension.
We observe the time dependence of the position distribution kurtosis 
and find
interesting behaviours. We point out a few
physical cases where the mass fluctuation problem   could be encountered as
a first approximation for agglomeration-fracture non equilibrium processes.
\end{abstract}


  \maketitle

\section{Introduction}

It is  well
known that, while the Maxwell-Boltzmann distribution takes place in any
system at  equilibrium, non-equilibrium systems present in general
qualitative and quantitative deviations from the former. A case of
particular interest is that of distributions characterized by a power
law tail, and therefore by an over-population for high energy as
compared to the Gaussian. Indeed, such behaviours occur in a large
number of physical systems, going from self-organized media to granular
gases, and may have striking consequences due to the large probability
of extreme/rare events.

The ubiquity of these {\em fat tail }
distributions in nature has motivated several attempts in the literature
to construct a general formalism for their description, one of the most
recent being the Tsallis thermodynamic formalism \cite{beck1,tsallis1}.
The latter is based on a proper
extremization of the non extensive entropy

\begin{equation}
\label{tsallis}
S_q =k \frac{(1-\sum_{i} p_i^q)}{q-1}
\end{equation}
that leads to generalized canonical
distributions, often called Tsallis distributions

\begin{equation}
\label{thermo}
f_q(x)=\frac{e_q^{-\beta^{'} x}}{Z}
\end{equation}
In this expression, $x$ denotes the state of the system, and $e_q^x$ is the
$q$-exponential function defined by

\begin{equation}
e_q^x \equiv (1+(1-q) x)^{\frac{1}{1-q}}
\end{equation}

This definition implies that $e_1^x = e^x$, and that the energy is
canonically distributed in the classical limit $q\rightarrow 1$, as
expected. Let us also note that  this formalism draws a direct
parallelism with the equilibrium theory, where $\beta^{'}$ plays the
role of the inverse of a temperature, and $Z$ that of a partition
function.

Tsallis distributions have been observed in numerous fields
\cite{tsallis2} but their fundamental origin is still debated due to
the lack of simple model and exact
treatment justifying the formalism.
  It is therefore important to study simple
statistical models in order to show in
which context Tsallis statistics apply.
There are several microscopic ways to justify Tsallis statistics. One of
them has been introduced some years ago by Beck\cite{beck1}.
Another has been recently introduced by Thurner\cite{thurner}.
Beck theory was first initiated by considering the
case of a Brownian particle moving in a specially thermally fluctuating medium,
i.e. the inverse temperature $\beta$ is a chi-squared distributed 
random variable.
Yet, Thurner \cite{thurner} has shown that one can derive Tsallis 
distributions in a general
way
without Beck ''{\em chi-square} assumption''.

In view of the above considerations,
the motion of a Brownian particle whose $mass$ is a random
variable seems to be a paradigmatic example. Moreover many
physical cases are concerned by such situations.
A sharp mass or volume variation of entities can be encountered in many
non-equilibrium cases: ion-ion reaction\cite{ionion1,ionion2,ionion3},
  electrodeposition\cite{EDP}, granular flow\cite{gold1, luding1, gran3}
   formation of
planets through dust aggregation\cite{dust1,dust2,dust3},
film deposition\cite{film}, traffic jams\cite{traffic1,traffic2,traffic3},
  and even stock markets\cite{GTA1,GTA2}
(in which the volume of exchanged shares fluctuates and the price
undergoes some random walk). It should be noticed that $a$ $priori$
fluctuating mass problems differ from temperature fluctuations. Two 
masses can be addded to
each other; this is hardly the case for temperatures. An expos\'e
of such a generalized Brownian motion
and the distinction between  $masses$ and $temperatures$ will be emphasized
  in Sect.II and Sect. III.

By supposing two time scales, i.e. assuming that relaxation processes
for the particle are faster than the characteristic times for the mass
fluctuation, it will be shown that its asymptotic velocity distribution
is in general non-Maxwellian. Moreover, for some  choice of
the mass probability distribution,  the Brownian particle
velocities are Tsallis distributed. We verify this result by performing
simulations of the corresponding Langevin equation.

Moreover, we also consider the case when  the mass fluctuation is {\em not}
adiabadically slow. We show in Sect. IV that the mass dependence of 
the relaxation
rate may have non-negligible consequences on the velocity distribution
of the Brownian particle. In order to fit the resulting
distributions, and to describe the deviations from Tsallis statistics,
we introduce truncated Tsallis distributions.
Finally, in Sect. V,
  we study the diffusive properties of the Brownian particle.
  To do so, we model the motion of the particle by a random walk
   with time-dependent jump probabilities, associated with the mass
   fluctuations of the particle. The model exhibits standard diffusion, i.e. $<x^2> \sim t$,
   but the shape of
   the scaling position distribution is anomalous and may exhibit quasi-stationary features.
Sect. VI serves as a summary and conclusions.

\section{Generalizing the Brownian motion to one in a fluctuating medium}

For setting the framework of the present study,
let us recall one microscopic way  to justify Tsallis statistics\cite{beck1}.
Beck theory, later
called {\em Superstatistics}, was first initiated by considering the
case of a Brownian particle moving in a fluctuating medium, i.e. an
ensemble of macroscopic particles evolving according to Langevin
dynamics

\begin{equation} \label{langevin} m \partial_t v = - \lambda v + \sigma
L(t) \end{equation}
where $\lambda$ is the friction coefficient, $\sigma$ describes the
strength of the noise, and $L(t)$ is a Gaussian white noise \footnote{From now
on, we restrict our analysis to the one-dimensional case for the sake of
writing simplicity, but without much loss of generality}.
  Contrary to the classical Brownian motion, however, one may consider that the
features of the medium may fluctuate temporally and/or spatially,
namely the quantity $\beta=\frac{\lambda}{\sigma^2}$, i.e. a quantity 
which plays the role of the
  inverse of
temperature, changes temporally on a time scale $\tau$, or on the
spatial scale $L$; see also \cite{luc}. E.g. in his original paper, 
Beck assumed that this
quantity fluctuates adiabatically slowly, namely that the time scale
$\tau$ is much larger than the relaxation time for reaching local
equilibrium. In that case, the stationary solution of the
non-equilibrium system consists in Boltzmann factors $e^{-\beta x}$ that
are averaged over the various fluctuating inverse temperatures $\beta$

\begin{equation} \label{beck} f_{Beck}(x) = \frac{1}{K} \int d\beta ~
g(\beta)~  e^{-\beta x} \end{equation}
where $K$ is a normalizing constant, and $g(\beta)$ is the probability
distribution of $\beta$. Let us stress that ordinary statistical
mechanics are recovered in the limit $g(\beta) \rightarrow \delta(\beta
- \beta_E)$. In contrast, different choices for the statistics of  $\beta$
may lead to a large variety of probability
distributions for the Brownian particle velocity.

Several forms for $g(\beta)$ have been studied in the literature
\cite{beck4}, but one  functional family of $g(\beta)$ is particularly
interesting. Indeed,  the generalized Langevin model Eq.(\ref{langevin})
generates Tsallis statistics for the velocities of the Brownian particle
if $\beta$ is a chi-squared random variable

\begin{equation} \label{chi2} g(\beta) = \frac{1}{b \Gamma(c)}
~(\frac{\beta}{b})^{c-1} ~e^{\frac{-\beta}{b}} \end{equation}
where $b$ and $c$ are positive real parameters which account for the
average and the variance of $\beta$. Let us stress that a chi-squared
distribution derives from the summation of squared Gaussian random
variables $X_i$,
$ \beta = \sum_{i=1}^{2c} X_i^2$,
where the $X_i$ are independent, and $<X_i>=0$. By introducing
Eq.(\ref{chi2}) into Eq.(\ref{beck}), it is straightforward to show that the
velocity distribution of the particle is Eq.(\ref{thermo}), if  one identifies
  $c=\frac{1}{q-1}$ and $bc =
\beta^{'}$.

For completeness, let us also mention  a study of Eq.(4) when
$\lambda$ fluctuates \cite{LTH}.

\section{Fluctuating mass}

Let us now consider the
  diffusive properties of a macroscopic cluster such as
one arising in granular media \cite{gold1,luding1} or in traffic 
\cite{traffic1}.
Such media are composed by a large number of macroscopic particles.
Due to their inelastic interactions, the systems are composed of
   very dense regions evolving
along  more dilute ones. In general, there is a continuous exchange of
particles between the dense cluster and the dilute region, so that the
total mass of the macroscopic entity is not conserved. As a first
approximation, we have  thus considered the simplest approximation for
this dynamics, namely the cluster is  one Brownian-like particle whose
mass fluctuates in the course of time. To mimic this effect, we have
assumed that (i) the distribution of masses is $a$ $priori$ given by
$g(m)$, and (ii) the mass of the cluster fluctuates with a
characteristic time $\tau$. By definition, this model evolves according
to the Langevin equation Eq.(\ref{langevin}),where $m$ is now the
random variable.

Given some realization
of the random mass, say $m=m_R$, one easily checks that the velocity
distribution of the particle converges toward the distribution

\begin{equation} f_{B}(v) \rightarrow \frac{\sqrt{\beta m_R
}}{\sqrt{2 \pi}} e^{-\beta \frac{m_R v^2}{2}} \end{equation}
This relaxation process takes place over a time scale $t_R\sim
\frac{m_R}{\lambda}$. Therefore, if the separation of time scales $t_R <<
\tau$ applies, the asymptotic velocity distribution of the cluster is
given by

\begin{equation} \label{limitS} f_{B}(v) = \int dm g(m)
\frac{\sqrt{\beta m }}{\sqrt{2 \pi}} e^{-\beta
\frac{m v^2}{2}}. \end{equation}
Consequently, this
leads to a Tsallis distribution  if the masses are
chi-squared distributed.
In that sense there is a direct correspondence between Beck approach and ours.
However there is more to see in the latter case because it justifies 
Tsallis non extensive
entropy approach in a more mechanistic way. Subsequently two basic
questions can be raised: (i) what are the limits of validity of such 
a non-equilibrium
approach?, and (ii) are the mass fluctuation time scales observable?

\section{Relaxation mechanisms}
\begin{figure}
  \includegraphics[angle=-90,width=3.5in]{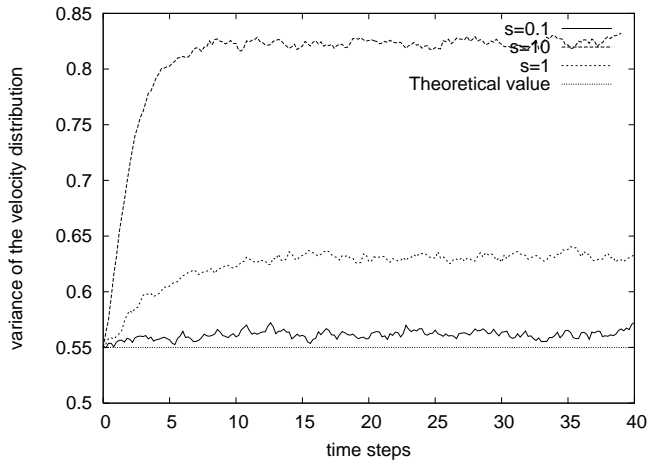}
  % Here is how to import EPS art
\caption{\label{figure1} Time evolution of the  $<v^2>$ 
for three values of
$\tau$, characterised by $s \equiv \frac{t_R}{\tau} = [0.1, 1, 10] $.  The possible masses in the system are $1$ and $10$.
Consequently, the  $<v^2>$ for each species are respectively $1.0$ and
$0.1$.  Moreover, in the slow limit ($\tau>>t_R$), the asymptotic velocity
fluctuation is, on average, $0.55$. When $\tau$ is small, most of the
particle velocities are distributed like the light ones, and their
energy is closer to $1.0$ than to $0.55$.}
\end{figure}

\begin{figure}
\includegraphics[angle=-90,width=3.5in]{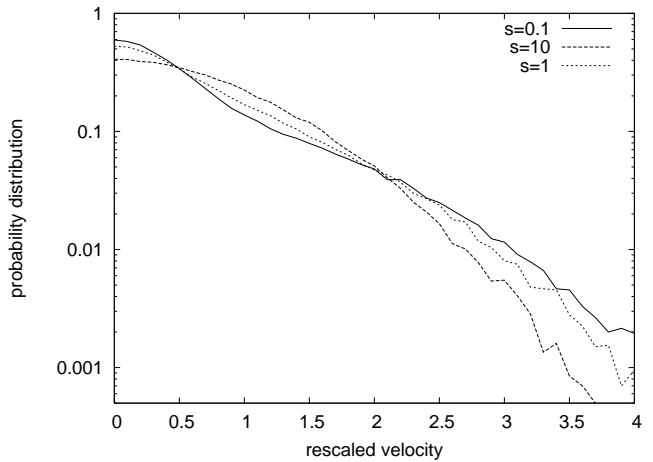}
\caption{\label{figure2} Asymptotic velocity distribution for
different relaxation time scales $s$ of a Brownian particle, namely $0.1$, $1$ and $10$.
The possible masses in the system are $1$ and $10$.
   The  velocities are rescaled so that $<v^2>=1$.}
  \end{figure}

In this section, we report numerical simulations of the random process
Eq.(\ref {langevin}). The objective is twofold. First, we verify the
theoretical prediction Eq.(\ref{limitS}) in the adiabatic limit. Next, this
allows to study systems which are beyond the range of validity of
Eq.(\ref{limitS}), namely systems where the separation of scales
$t_R<<\tau$ does not apply, thereby investigating the effects of
competition between the relaxation to equilibrium and the fluctuating
features of this equilibrium state. This program is achieved by
considering  three different relaxation characteristic time scales for
the processes, namely $s=\frac{t_R}{\tau} = (0.1, 1.0, 10)$. Moreover,
we first consider a paradigmatic case when the masses can switch between
two different discrete values, each with equal probability:
\begin{equation} 
\label{double} 
g(m) =\frac{1}{2} (\delta(m-m_1)
+\delta(m-m_2)) 
\end{equation} 
If Eq.(\ref{limitS}) applies, i.e. in the slow 
fluctuation limit,
  Eq.(\ref{double}) leads to

\begin{equation} \label{limitS2} f_{B}(v) = \frac{1}{2}
\sqrt{\frac{\beta}{ 2 \pi}} (\sqrt{m_1}
e^{-\beta \frac{m_1 v^2}{2}} + \sqrt{m_2}
e^{-\beta \frac{m_2 v^2}{2}}). \end{equation}

\begin{figure}
  \includegraphics[angle=-90,width=3.5in]{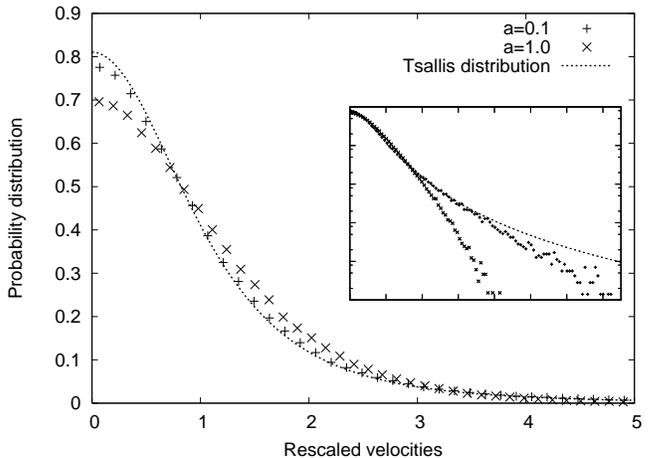}
\caption{\label{figure3} Truncated Tsallis distribution 
$ g_2(m)$ for
$a=0.1$, $a=1.0$ and $b=\infty$ in rescaled velocities, so that
$<|v|>=1$. The solid line is the Tsallis distribution
$\frac{8}{\pi^2}\frac{1}{(1+ (\frac{2}{\pi})^2 v^2)^2}$. The cut-off
procedure gives more weight to the central range of the distribution
($m\in[1,3]$), and decreases the importance of the peak and of the 
tail. The inset
corresponds to these functions in log-normal scale, in the interval   $[0,15]$.
}
\end{figure}

Before focusing on the velocity distributions for the Brownian
particles, let us stress that their average energy depends on the speed
of the fluctuation mechanism. Indeed, in the slow limit $s<<1$, the
energy of the cluster converges very rapidly toward the equipartition
value $m_i <v^2> = e$, where $e$ is the average kinetic energy of the
bath and $m_i $ is the mass of the cluster at that time. This implies
that the fluctuations of the measured velocities are given by
$<v^2>=\frac{1}{2}(\frac{e}{m_1}+\frac{e}{m_2})$. In contrast, in the
faster limit $s>1$, the dependence of the characteristic time $t_R\sim
\frac{m}{\lambda}$ can not be neglected. Indeed, this relation implies
that particles with a smaller mass relax faster than particles with a
larger mass. Consequently, the lighter particles should  have a value
close to their equipartition value $m_i <v^2> = e$, while the heavier
particles should have an energy larger than their expected value. This
property is verified (Fig. \ref{figure1}).

In order to compare the velocity distributions for different time scales
$\tau$, and therefore at different energies, we rescale the velocities
so that $<v^2>=1$. The results, as plotted in Fig. \ref{figure2},
confirm the theoretical predictions Eq.(\ref{limitS2}) and our 
description in the previous
paragraph. Indeed, in the slow limit $s=0.1$, the velocity distribution
converges toward the leptokurtic distribution Eq.(\ref{limitS2}), i.e. a
distribution with a positive kurtosis and an overpopulated tail. In
contrast, when the mass dependence of the $t_R$ has to be taken into
account, most of the particles have velocities distributed like those of
the light particles, namely are Maxwell-Boltzmann distributed.

\begin{figure}
\includegraphics[angle=-90,width=3.5in]{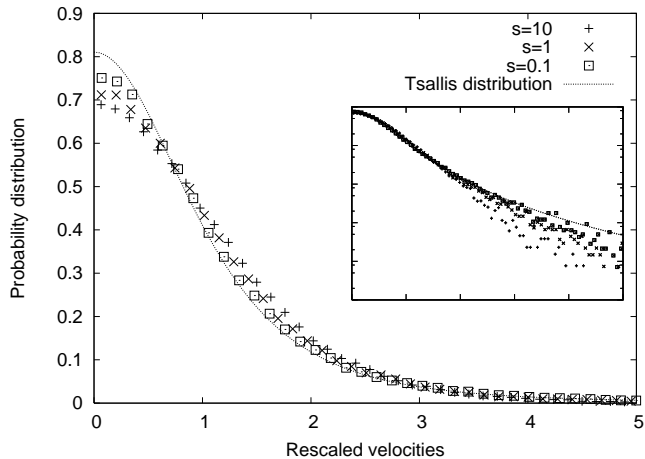}
\caption{\label{figure3} Asymptotic solutions of the 
Langevin equation
for 3 time scales, where we fix the velocity scale  $<|v|>=1$. The solid
line is the theoretical distribution  obtained from (\ref{beck}).  }
\end{figure}

\begin{figure}
\includegraphics[angle=-90,width=3.5in]{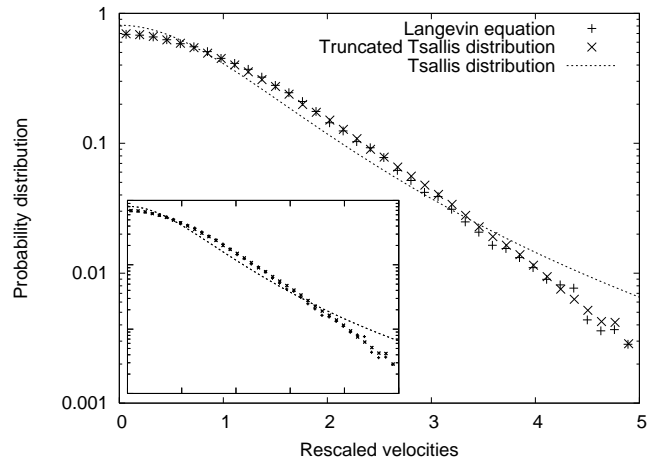}
\caption{\label{figure4} Stationary solution of the 
Langevin equation
for $s=10$ and truncated Tsallis distribution for $a=1$. The dotted line
is the Tsallis distribution $\frac{8}{\pi^2}\frac{1}{(1+
(\frac{2}{\pi})^2 v^2)^2}$. }
\end{figure}

In the case of general Tsallis distributions, obtained through
chi-squared distributed cluster masses, the problem is more complex due
to the continuum of masses in the system, and  to the associated
continuum of characteristic relaxation times.  Moreover, the existence
of extreme values for the masses may cause non-realistic numerical
problems. Indeed, arbitrary small masses lead to arbitrary high values
of the velocities. In the following, we avoid this effect that is responsible
for the power law
tails observed in Tsallis distributions. This is justified by the fact that
any physical system has a minimum size for its
internal components. Similarly, we restrict the maximum size of the
clusters in order to avoid infinitely slow relaxation 
processes. These limitations
are formalized by using the following truncation for the mass
distributions, inspired by truncated L\'evy distributions
\cite{mantegna1}

\begin{eqnarray} \label{truncated} g(m; c)  &=& k ~ \chi^2(m; c) ~~
\mbox{if} ~ m>  a ~ \mbox{and}~ m< b\cr g(m; c)  &=& 0 ~~
\mbox{otherwise} \end{eqnarray}
where $c$ is a parameter characterizing
the chi-squared distribution,  $k$ is a normalizing constant, and 
$a<b$ are cut-off parameters. In the
following, we use $a=0.01$ and $b=100.0$ in the simulations of the
Langevin equation. As shown in Fig. \ref{figure3}, this procedure is a
natural way to smoothen the tail of the Tsallis velocity distribution
while preserving its core. This method, that will be discussed further
in a forthcoming paper, should be applicable to  large variety of
problems (like those mentioned in the introduction) where extreme events
have to be truncated for physical reasons.

Numerical simulations of the Langevin equation for chi-squared mass
distributions generalize in a straightforward way the results obtained
from the 2-level distribution Eq.(\ref{double}). Indeed, the faster the
mass distribution fluctuates the larger deviations from the Tsallis
distribution are. Moreover, these deviations have a tendency to
underpopulate the tail of the distribution, and therefore to avoid the
realization of extreme values of the random process.

It is also
important to note the similarities between these asymptotic solutions of
the Langevin equation and the truncated Tsallis distributions defined by
Eq.(\ref{truncated}). They highlight the flexibility of truncated Tsallis
distributions in order to describe deviations from the Tsallis
distributions, as shown in Fig. \ref{figure4}.

\section{Diffusion}
In this last section, we focus on the diffusive properties of a 
Brownian particle
with fluctuating mass. Therefore, we take into account the hydrodynamic time
scale $t_H$, that is associated to the evolution of spatial inhomogeneities.
  It is well known \cite{resi}, when the mass of the Brownian particle 
is constant in time,
   that separation of time scales $t_H >> t_R $ is required in order to
derive the diffusion equation:
\begin{equation}
\label{diffusion}
\partial_t n({\bf x}; t) = D \partial_{x^2} n({\bf x}; t)
\end{equation}
where $D$ is fixed by the Einstein formula.

\begin{figure}
\includegraphics[angle=-90,width=3.50in]{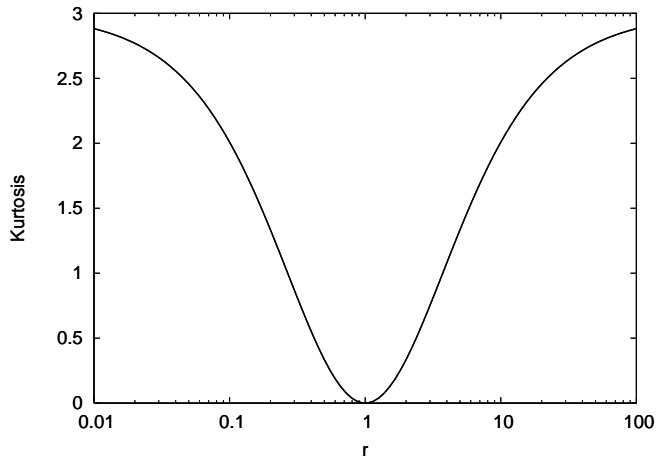}
\caption{\label{relation} Kurtosis of the distance
distribution Eq.(\ref{kuku}), for two Bernouilli random walkers
in one dimension with different walk lengths. }
\end{figure}

When the mass of the Brownian particle fluctuates, however, we have
shown above that there is an additional time scale $\tau$ in the dynamics.
As first approximation, we
restrict the scope to the limit of very slow fluctuations ($\tau>> t_H$),
that is more restrictive than the limit discussed in the previous 
section ($\tau>> t_R$).
In that case, it is possible to show  \cite{creta} that
the Chapman-Enskog procedure leads to Eq.(\ref{diffusion}),
where $D$ is now time-dependent, i.e. a random variable that is a 
function of the mass of
the Brownian particle.

In the following, we investigate the process associated to Eq.(\ref{diffusion})
  with fluctuating diffusion coefficient. To do so, we simplify the 
analysis by considering
  a one-dimensional discrete time random walk, where the jump 
probabilities may fluctuate
  in time \cite{fluct}. Namely, the walker located at $x$ performs at 
each time step
   a jump of length $l$, with Bernouilli probabilities:
\begin{equation}
\label{bernouilli}
P(k)|_l = \frac{1}{2} [\delta(k,l) + \delta(k,-l)]
\end{equation}
The quantity $l$ fluctuates between two integer values $l_A\leq l_B$ 
that correspond
to an heavy/light state for the Brownian particle. When $l_A=l_B$, it 
is easy to show
that the first distance moments $m_i=<x^i >$ asymptotically behave like:
\begin{eqnarray}
\label{asym}
m_2&=&D t \cr
m_4&=&3 D^2 t^2
\end{eqnarray}
where the diffusion coefficient $D=l_A^2$.
Relation (\ref{asym}) implies that the kurtosis
  of the distance distribution $\kappa=\frac{m_4}{m_2^2}-3$ vanishes
  asymptotically, as required by the central-limit theorem.

\begin{figure}
\includegraphics[angle=-90,width=3.50in]{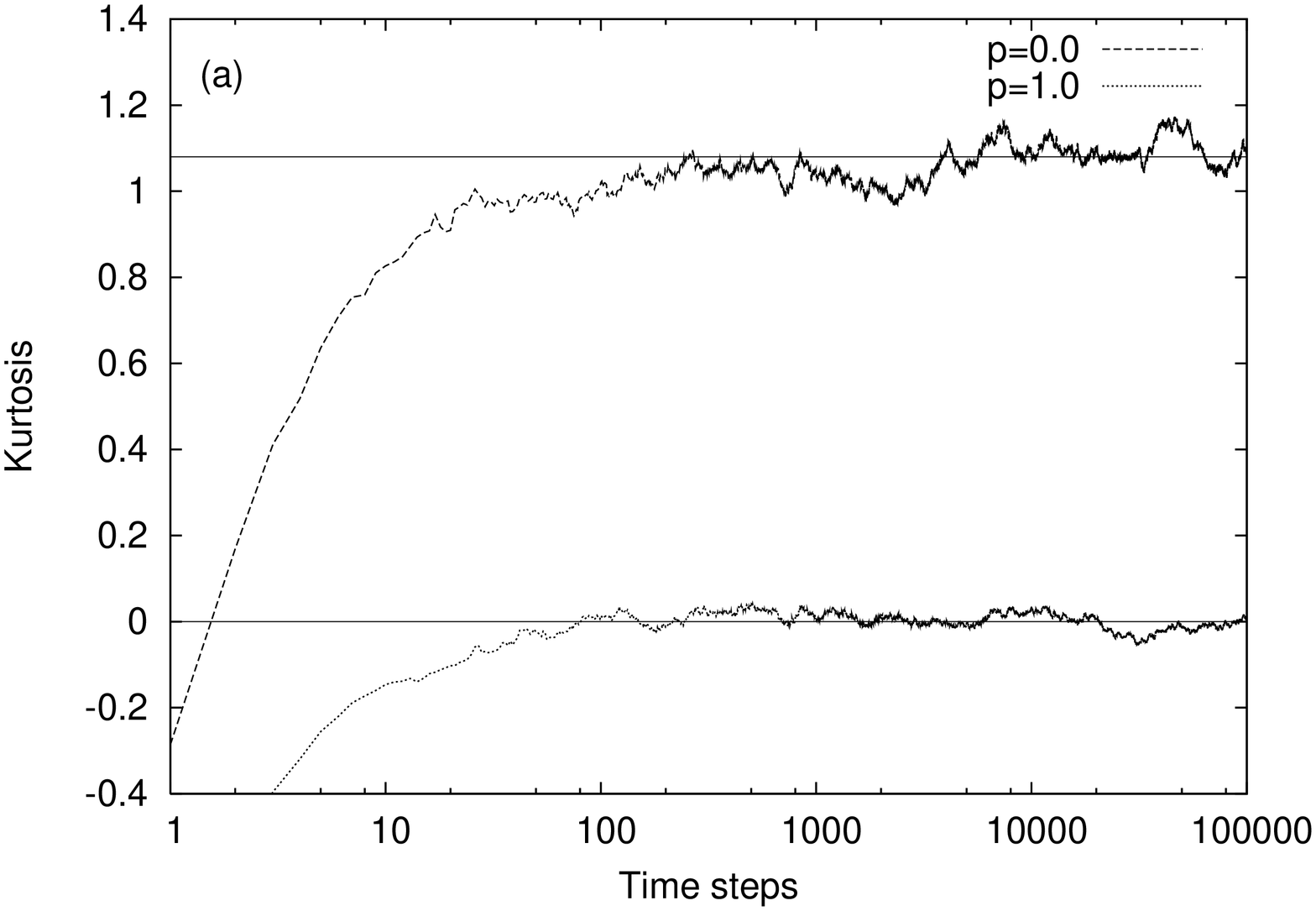}

\includegraphics[angle=-90,width=3.50in]{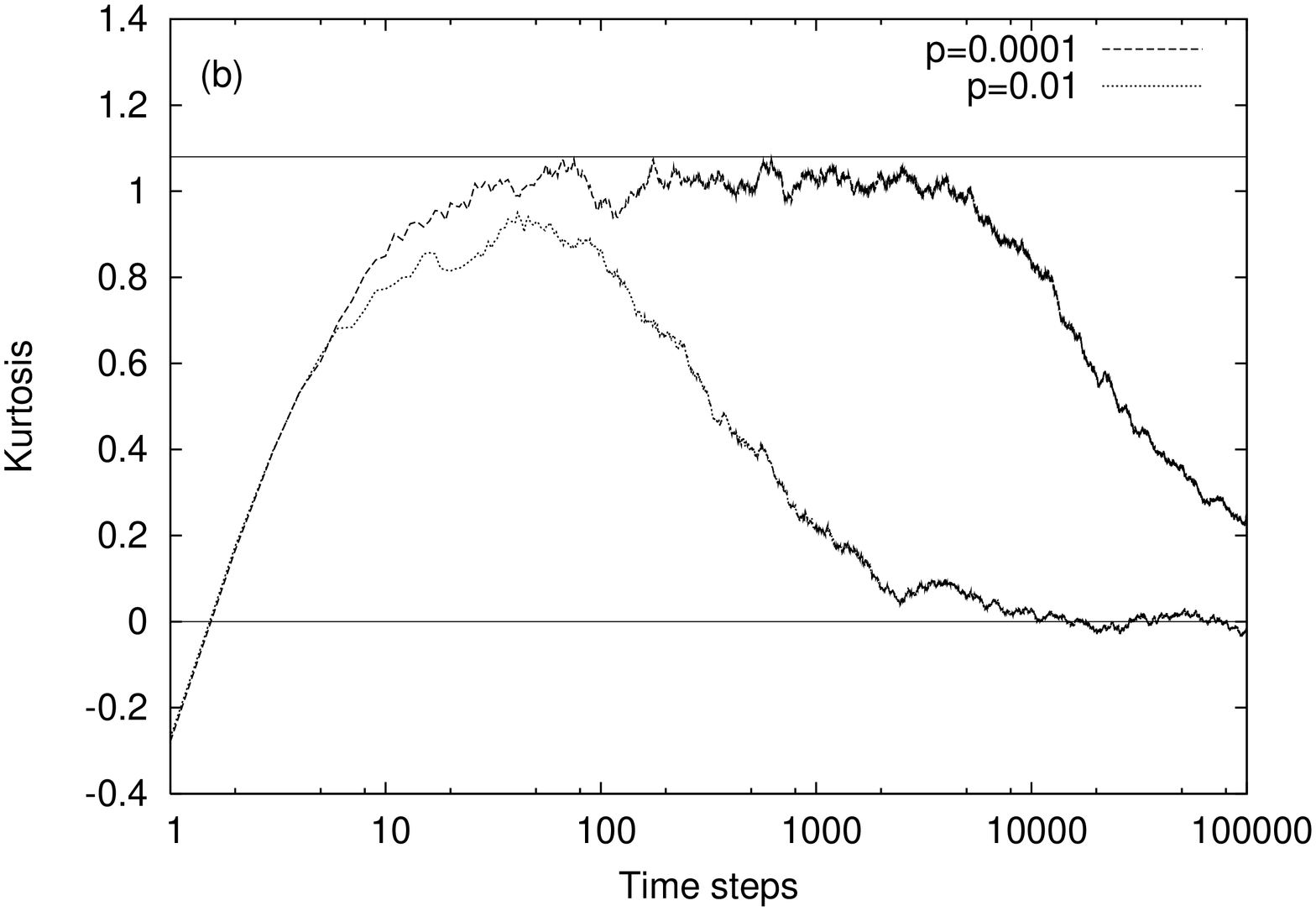}
\caption{\label{diffusion} Kurtosis $\kappa$ of the 
distance distribution,
as a function of time, for the Bernouilli random walker with fluctuating mass
(see main text). In (a), we focus on the limiting cases $p=0.0$ and 
$p=1.0$ that
  converge  toward the theoretical prediction, Eq.(\ref{kuku}),
  $\kappa=1.08$ and $\kappa_{G}=0$ (Gaussian) respectively.
   These asymptotic values are represented by solid lines.  In (b), we 
focus on the
    intermediate cases $p=0.01$ and $p=0.0001$. The state associated 
to Eq.(\ref{kuku})
    is stationary during a long time extent,
    and is followed by a convergence toward the Gaussian.}
\end{figure}

Another simple limit consists in a system with $l_A\neq l_B$, and where the
  mass fluctuation process is infinitely slow. Consequently, the 
particles do not
   change mass and the system is composed of two species that diffuse 
differently.
   In that case,  the first distance moments read:
\begin{eqnarray}
\label{asym2}
m_2&=&\frac{1}{2} (D_A+D_B) t \cr
m_4&=&\frac{3}{2} (D_A^2+D_B^2) t^2
\end{eqnarray}
where the diffusion coefficients are $D_A=l_A^2$ and $D_B=l_B^2$.
This relation implies that the asymptotic kurtosis is equal to:
\begin{equation}
\label{kuku}
\kappa = 6 \frac{(D_A^2+D_B^2)}{(D_A+D_B)^2} -3
\end{equation}
In Fig.\ref{relation}, we plot $\kappa(r)= 6 \frac{(1+r^2)}{(1+r)^2} -3$,
  where $r=\frac{D_A}{D_B}$. The figure shows that $\kappa \geq 0$, i.e.
  the distribution is characterised by an overpopulated tail, except in the
  usual limit $r=1$. This result is expected, as a system with $r \neq 1$ is
   composed of two species that explore the space at different speeds. Let us stress
   that despite this anomalous position distribution, the diffusion is standard, i.e. $<x^2> \sim t $ (see Eq.\ref{asym2}).

In order to study intermediate situations, we have  performed 
computer simulations
of the random walk. The system is composed of $50000$ walkers, that 
are initially
  located at $x=0$ and randomly divided in the species $A$/$B$.
The mass fluctuations are uncorrelated and occur with probability $p 
\in{[0,1]}$ at
  each time step, i.e. the characteristic time of the fluctuations is 
$\tau \sim p^{-1}$.
In the simulations, we have used $l_A=1$ and $l_B=2$, so that
the prediction for the kurtosis in the limit $\tau \rightarrow 
\infty$ is $\kappa_{AB}=1.08$.
This prediction (Fig.\ref{diffusion}a) is verified by simulations in 
a system where
the mass of a walker is constant in time.
In the limit $\tau=1$, i.e. the walker changes mass at each time step,
  the kurtosis converges rapidly toward the gaussian value 
$\kappa_G=0$. In contrast,
   for higher finite values of $\tau$ (Fig.\ref{diffusion}b), one observes a
crossover between the two asymptotic behaviours $\kappa_{AB}$ and $\kappa_{G}$.
  The stability of the state $\kappa_{AB}$ is longer and longer for 
increasing values of $\tau$.
This quasi-stationarity \cite{lambi} originates from the following reason.
  Over a long time period $T$, with $T>>\tau$, the particles have 
suffered so many changes
   from state $A$ to $B$ and forth, that their asymptotic dynamics is 
equivalent to that of
    a random walk with 2 possible jump lengths $l_A$ and $l_B$, with 
probabilities:
\begin{equation}
\label{bernouilli2}
P(k) = \frac{1}{4} [\delta(k,l_A) + \delta(k,-l_A)+\delta(k,l_B) + 
\delta(k,-l_B)]
\end{equation}
Consequently, its asymptotic dynamics is subject to the classical
  central-limit theorem, and the position distribution is a Gaussian 
for $T>>\tau$.

\section{Conclusions}

In examining an appartently unusual generalization of the old
Brownian motion problem, i.e. a particle with fluctuating mass,
we have found a very simple example justifying
  non-extensive thermodynamics.
However the simplicity is related to underlying considerations
on quite various physical (or other) systems in which some ''mass''
is evolving with time, sometimes stochastically, as mentioned in the 
introduction of this paper.

For the sake of generality, we have studied systems with arbitrary 
mass distribution and time scale
  of the mass fluctuation. Let us stress that the mass statistics has 
been chosen $a$ $priori$,
  and that a more detailed study requires dynamical treatment of the
fluctuating mass by Langevin equation. This additional modelling 
depends on the nature
  of the considered problem.
The velocity distribution of the Brownian particle has been studied 
by performing
simulations that highlight the important role of mass fluctuation time scale.
In the case of chi-squared mass distribution, it is shown that {\em truncated}
Tsallis distributions seem to describe in a relevant way deviations 
from the Tsallis
statistics. Such distributions should be applicable to a large variety of
problems where extreme events
have to be truncated for physical reasons, e.g. finite size effects, 
- when there is no infinity! Among these possible applications, let us note their
occurrence in airline disasters statistics \cite{lambi2}.

We have also studied the diffusive properties of the Brownian particle.
  In the limit of slow mass fluctuation times, the particle motion is 
modelled by
  a random walk with time-dependent jump probabilities. Moreover, for 
the sake of clarity,
  we restrict the scope to a dichotomous mass distribution. This 
modelling is a simplification of
   the complete problem, that has to be justified by a complete 
analysis starting
   from the Fokker-Planck equation itself \cite{lambi2}.  Nonetheless, despite
   its apparent simplicity, the random walk analysis shows non-trivial 
behaviours, namely the system is characterised by standard diffusion, 
   associated with non-Gaussian scaling distributions and quasi-stationary solutions.
    These features originate from very general mechanisms that suggest 
their relevance
    in various systems with  fluctuating "mass" parameters.

    In conclusion, we have generalized Brownian motion to the case of 
fluctuating mass systems, and
    distinguish them from systems in which there are local temperature 
fluctuations. We have found
that the velocity of such a particle may exhibit various anomalous distributions,
including Tsallis distributions and truncated Tsallis distributions. The 
study of a one dimensional
random walk also indicates  anomalies 
in the kurtosis of
the position distributions which might indicate physical biases in various 
processes as those
recalled in the introduction. This suggests to look for the value of 
higher order
distribution moments and on their time evolution for understanding properties
of the non-equillibrium systems.

\begin{acknowledgments}
  R.L. would like to thank ARC 02-07/293 for financial support. M.A 
thanks A. Pekalski
  for comments and suggestions.
\end{acknowledgments}

\end{document}